\documentclass[twocolumn,amsmath,amssymb]{revtex4}
\usepackage{graphicx}
\usepackage{dcolumn}
\usepackage{color}
\usepackage{scrextend}
\usepackage{subfig}
\usepackage{bm}
\usepackage{amsmath,amssymb,graphicx}
\usepackage{epsfig}
\usepackage{enumerate}
\usepackage{array}
\usepackage{multirow}

\begin{document}
\title
{Energetics of the dissipative quantum oscillator}
\author{Aritra Ghosh\footnote{ag34@iitbbs.ac.in}, Jasleen Kaur\footnote{jk14@iitbbs.ac.in}, and Malay Bandyopadhyay\footnote{malay@iitbbs.ac.in}}
\affiliation{School of Basic Sciences,\\ Indian Institute of Technology Bhubaneswar, Argul, Jatni, Khurda, Odisha 752050, India}
\vskip-2.8cm
\date{\today}
\vskip-0.9cm

\textit{Dedicated to Prof Sushanta Dattagupta, on the\\ occasion of his 75th birthday with deep respect and admiration}
\vspace{5mm}


\begin{abstract}
In this paper, we discuss some aspects of the energetics of a quantum Brownian particle placed in a harmonic trap, also known as the dissipative quantum oscillator. Based on the fluctuation-dissipation theorem, we analyze two distinct notions of thermally-averaged energy that can be ascribed to the oscillator. These energy functions, respectively dubbed hereafter as the mean energy and the internal energy, are found to be unequal for arbitrary system-bath coupling strength, when the bath spectral function has a finite cutoff frequency, as in the case of a Drude bath. Remarkably, both the energy functions satisfy the quantum counterpart of the energy equipartition theorem, but with different probability distribution functions on the frequency domain of the heat bath. Moreover, the Gibbs approach to thermodynamics provides us with yet another thermally-averaged energy function. In the weak-coupling limit, all the above-mentioned energy expressions reduce to \(\epsilon = \frac{\hbar \omega_0}{2} \coth \big(\frac{ \hbar \omega_0}{2 k_B T}\big)\), which is the familiar result. We generalize our analysis to the case of the three-dimensional dissipative magneto-oscillator, i.e., when a charged dissipative oscillator is placed in a spatially-uniform magnetic field.
\end{abstract}

\maketitle

\section{Introduction}
In recent years, extensive studies have been found to highlight the critical role that a dissipative environment plays in mesoscopic systems, fundamental quantum physics, and quantum computation \cite{a,b,c,d}. This has opened the doorway to a rigorous investigation of many results that were derived for macroscopic systems. In particular, one may mention substantial investigations in the area of quantum and mesoscopic thermodynamics \cite{e,f}. In some cases, the validity of the fundamental laws and relations related to quantum thermodynamics have been questioned, particularly at low temperatures where quantum effects are predominant \cite{g}. Further, several interesting new facets of old results have come out while exercising the validity of the fundamental laws or definitions. Since many subtle issues may arise, one needs to be cautious while questioning the validity of the fundamental laws (especially the fundamental laws and definitions of thermodynamics).\\

In this context, a harmonic oscillator coupled to a heat bath can be a good starting point. Such a model plays a crucial role in physics, like modeling an atom coupled to a heat bath, chemical reactions in a molecular environment, an atom in a blackbody radiation (BBR) field, accelerated electron in a BBR field, and a magnetic flux trapped in a current-biased Josephson junction or in the superconducting ring of a superconducting quantum interference device (SQUID) \cite{h}. Thus, the problem of a quantum harmonic oscillator which is interacting with an environment is a rather interesting one. Our purpose here is to analyze different fundamental notions of (thermal) energy associated with a dissipative oscillator under an `independent-oscillator' model of a heat bath \cite{lho1}. In such a model, the environment, i.e., the heat bath, is taken to be composed of an infinite number of independent quantum oscillators (see also \cite{FV, CL}).\\

It is well known from the classical equipartition theorem that the thermal energy of a classical one-dimensional oscillator is \(k_B T\), whereas, for the quantum mechanical case, it reads
\begin{equation}\label{11111}
\epsilon(\omega_0,T) = \frac{\hbar \omega_0}{2} \coth \bigg(\frac{\hbar \omega_0}{2 k_B T}\bigg).
\end{equation}
These results are obtained based on the assumption that the equilibrium properties of the system (here, the oscillator) can be described by the canonical partition function \cite{R}, which reads
\begin{equation}
Z = \sum_{n = 0}^\infty e^{-\beta \epsilon_n}, \hspace{6mm} \epsilon_n = \hbar \omega_0 \bigg(n + \frac{1}{2}\bigg),
\end{equation} for the quantum mechanical case, while for the classical case, one replaces the sum with an integral on the phase space. The partition function is obtained by letting the system stay in contact with a heat bath of temperature \(T = 1/k_B \beta\), under two important assumptions: (a) The heat bath by definition, is an infinite reservoir of energy, i.e., it has infinite specific heat, (b) the coupling between the system and the heat bath is sufficiently weak, such that the number of microstates describing the system and the heat bath taken (interacting) together is to a good approximation, just the product of the number of microstates of the system and the heat bath taken individually (without interactions) \cite{R}. Moreover, we take the spectrum of the oscillator to be \(\epsilon_n = \hbar \omega_0 (n + 1/2)\), which is true only for a free oscillator and not for one that is interacting with a heat bath. Thus, Eq. (\ref{11111}) can be justified only in the weak-coupling limit, while in a general case, there are bath-induced broadening effects on the energy levels. \\

In general, the system and the heat bath may get coupled with an arbitrary strength. In such a situation, Eq. (\ref{11111}) is no longer true and one has to resort to a more general framework to find the thermally-averaged energy of the oscillator. In any case, the starting point is to realize that the Hamiltonian which describes the system, the heat bath, and their interactions, has the generic structure: \(H = H_S + H_B + H_{SB}\), where the subscript `\(S\)' denotes system and `\(B\)' indicates the heat bath. A commonly used prototype is the independent-oscillator model \cite{lho1,FordQT0,purisdgbook} (see also \cite{FV,CL,QLE}), for which the heat bath is taken to be composed of an infinite number of independent quantum oscillators, such that the total Hamiltonian reads as
\begin{equation}\label{Hmodel}
H = \frac{p^2}{2m} + \frac{m \omega_0^2 x^2}{2}  + \sum_{j = 1}^N \Bigg[ \frac{p_j^2}{2m_j} + \frac{m_j \omega_j^2}{2} \big(q_j - x \big)^2 \Bigg].
\end{equation} Here, \(x\) and \(p\) are the position and momentum operators of the system, i.e., \([x,p] = i \hbar\), while \(q_j\) and \(p_j\) are the position and momentum operators of the \(j\)-th oscillator of the heat bath, i.e., \([q_j,p_k] = i \hbar \delta_{jk}\). At thermal equilibrium, the state of the system and the bath, including their interactions, all taken together, is described by the canonical distribution:
\begin{equation}\label{rhoH}
\rho_H = \frac{e^{-H/k_B T}}{Z}.
\end{equation}
It turns out that one is only interested in the thermodynamics of the system, while it is interacting with the bath, but not the one described by Eq. (\ref{rhoH}). This requires one to find the reduced density matrix of the system, call it \(\rho\), which involves computation of the so-called influence functional for tracing out the bath degrees of freedom \cite{FV,Ingold}. Except for the weak-coupling limit, the reduced distribution \(\rho\) does not coincide with \(\rho_{H_S} = e^{-H_S/k_B T}/Z_S\), where \(H_S\) is free-system Hamiltonian, i.e., the first two terms of Eq. (\ref{Hmodel}). \\

 The purpose of the present paper is to discuss and assess the different notions of thermally-averaged energy that can be associated with the dissipative quantum oscillator. In particular, we describe two distinct and unequal perceptions of energy that can be ascribed to the system based on the fluctuation-dissipation theorem, while performing a comparison of the two. These two energy functions are respectively dubbed as the mean energy and the internal energy. Remarkably, it is found that both the energy functions respect the quantum counterpart of the energy equipartition theorem, which has generated a considerable amount of interest in recent times \cite{jarzy1,jarzy2,jarzy3,jarzy4,jarzy45,jarzy5,jarzy6,kaur,kaur1,kaur2,kaur3}. We also compute the thermally-averaged energy from the partition function of the system and compare the result with the other two energy functions obtained from the fluctuation-dissipation theorem. For the one-dimensional dissipative oscillator, the internal energy proposed in \cite{FordQT0, FordQT, bb} is demonstrated to (approximately) coincide with the thermally-averaged energy obtained from the partition function. Following this, we extend our analysis to the case of the three-dimensional dissipative oscillator in a magnetic field, the so-called dissipative magneto-oscillator (see \cite{kaur,kaur3,molu} and references therein). Our analysis will help to clarify the similarities and differences between the above-mentioned energy functions. Besides comparing results obtained from different definitions of energy functions,  we can contrast the distinct ways in which they are defined. We express the internal energy function in the form of infinite series for both the one-dimensional dissipative oscillator and the three-dimensional dissipative magneto-oscillator, which, to the best of our knowledge are both novel expressions that have not appeared in the literature before.\\

The rest of the paper is organized as follows. In the next section [Sec. (\ref{FEsec})], we discuss the energetics of the one-dimensional dissipative quantum oscillator, analyzing the behavior of two distinct and unequal energy functions, while also comparing and contrasting the two. We discuss different limiting cases of the results and also describe the quantum counterpart of the energy equipartition theorem. For comparison, we also present the thermally-averaged energy function as obtained directly from the partition function of the system \cite{Ingold}. Then, in Sec. (\ref{magSec}), we generalize the results for the three-dimensional dissipative magneto-oscillator. We conclude the paper with some discussion in Sec. (\ref{DSec}).

\section{Energy of a one-dimensional dissipative quantum oscillator}\label{FEsec}
We now describe two distinct thermally-averaged energy functions that can be ascribed to the Brownian oscillator. These are termed respectively, the mean energy function and the internal energy function. We hope this section will clarify the differences between the two, which seem to have been unnoticed in the literature. Moreover, as we shall show, there are various similarities between the two. \\

Let us introduce the notion of susceptibility. The reduced dynamics of the system is described by the quantum Langevin equation [see App. (\ref{PreSec}) for details] \cite{lho1,purisdgbook,QLE}:
\begin{equation}\label{qle}
m \ddot{x}(t) + \int_{-\infty}^t \mu(t - t') \dot{x}(t') dt' + m \omega_0^2 x(t) = F(t),
\end{equation} which is a linear integro-differential equation for the system's position operator \(x = x(t)\). We now state without proof, the Callen-Welton fluctuation-dissipation theorem, in the form as applicable to the system under consideration. Consider a general perturbing force \(f(t)\) acting on the system. Then, one can solve the quantum Langevin equation via a Fourier transform as
\begin{equation}
\tilde{x}(\omega) = \alpha^{(0)}(\omega) \big[\tilde{F}(\omega) + \tilde{f}(\omega) \big] ,
\end{equation} where \(\alpha^{(0)} (\omega)\) is called the generalized susceptibility, given by
\begin{equation}\label{alphadef11}
\alpha^{(0)}(\omega) = \frac{1}{m(\omega_0^2 - \omega^2) - i \omega \tilde{\mu}(\omega)}.
\end{equation}
We denote by a `tilde', the Fourier transform of a quantity. Then, the fluctuation-dissipation theorem states that \cite{callenwelton,Kubo,case,bb}
\begin{widetext}
\begin{eqnarray}
\frac{1}{2} \langle x(t) x(t') + x(t') x(t) \rangle = \frac{\hbar}{\pi} \int_0^\infty {\rm Im} [\alpha^{(0)}(\omega + i 0^+)]  \coth \bigg(\frac{\hbar \omega}{2k_B T}\bigg) \cos [\omega(t-t')] d\omega,  \label{CW}
\end{eqnarray}independent of \(f(t)\) (the perturbing force is small). One may find the velocity autocorrelation function by differentiating Eq. (\ref{CW}) with respect to \(t\) and \(t'\), giving
\begin{eqnarray}
\frac{1}{2} \langle \dot{x}(t) \dot{x}(t') + \dot{x}(t') \dot{x}(t) \rangle = \frac{\hbar}{\pi} \int_0^\infty \omega^2 {\rm Im} [\alpha^{(0)}(\omega + i 0^+)]  \coth \bigg(\frac{\hbar \omega}{2k_B T}\bigg) \cos [\omega(t-t')] d\omega.  \label{CW1}
\end{eqnarray}
\end{widetext}
One may also compute correlation functions involving the bath variables such as \(\frac{1}{2}\langle q_j(t) q_k(t') + q_k(t') q_j(t) \rangle\), \(\frac{1}{2}\langle q_j(t) x(t') + x(t') q_j(t) \rangle\), and so on using different versions of the fluctuation-dissipation theorem which involve susceptibilities different from \(\alpha^{(0)}(\omega)\). Here, we do not pursue this explicitly.

\subsection{Mean energy function}\label{FEsec2}
Let us describe what we shall call the mean energy function of the dissipative oscillator. Consider the system Hamiltonian: \(H_S = \frac{p^2}{2m} + \frac{m \omega_0^2 x^2}{2}\), where \(x(t)\) and \(p(t)\) are now determined by the quantum Langevin equation [Eq. (\ref{qle})]. Let us take the average of the system Hamiltonian, i.e., consider
\begin{equation}\label{HSAvg}
\langle H_S \rangle = \frac{m \langle \dot{x}^2\rangle }{2} + \frac{m \omega_0^2 \langle x^2 \rangle}{2} ,
\end{equation}
where \(\langle \cdot \rangle\) denotes averaging over the noise ensemble and we have used the Heisenberg equation \(\dot{x} = \frac{[x,H]}{i\hbar} = \frac{p}{m}\). We remind the reader that although \(H_S\) describes the Hamiltonian of the system alone, the full dynamics is described by Eq. (\ref{Hmodel}), i.e., although we are averaging over \(H_S\) in Eq. (\ref{HSAvg}), this takes into account the bath-induced effects on the variables \(x\) and \(p\), as explicit in the quantum Langevin equation.\\

In order to obtain the equal-time averages \(\langle x^2 \rangle\) and \(\langle \dot{x}^2 \rangle\), we use the fluctuation-dissipation theorem mentioned in Eq. (\ref{CW}). The average \(\langle x^2 \rangle\) is obtained by setting \(t=t'\) in Eq. (\ref{CW}), while \(\langle \dot{x}^2\rangle\) is obtained by differentiating Eq. (\ref{CW}) once with respect to \(t\), and then with respect to \(t'\) before finally setting \(t = t'\). Combining these with Eq. (\ref{HSAvg}), one obtains
\begin{equation}\label{Edef111}
E := \langle H_S \rangle = \frac{m}{\pi} \int_0^\infty \epsilon(\omega,T) {\rm Im} [\alpha^{(0)}(\omega + i 0^+)] \bigg(1 + \frac{\omega_0^2}{\omega^2}\bigg) \omega d\omega,
\end{equation}
where \(\epsilon(\omega,T) := \frac{\hbar \omega}{2} \coth \big(\frac{\hbar \omega}{2k_B T}\big)\). Performing this integral, one can find the mean energy of the system.\\

Now, using the fact that \(\epsilon(\omega,T)\) is an even function of \(\omega\), we can write
\begin{widetext}
\begin{equation}\label{Edef111112}
E(T) = \frac{1}{2 \pi} \int_{-\infty}^\infty  \frac{\epsilon(\omega,T) (\omega^2 + \omega_0^2) {\rm Re}[\tilde{\gamma}(\omega)]}{\big[ (\omega_0^2 - \omega^2 + \omega {\rm Im}[\tilde{\gamma}(\omega)] \big]^2 + \big(\omega {\rm Re}[\tilde{\gamma}(\omega)]\big)^2} d\omega,
\end{equation}
where \(\tilde{\gamma}(\omega) = \tilde{\mu}(\omega)/m\), and we have also used the fact that the real and imaginary parts of \(\tilde{\gamma}(\omega)\) are respectively, even and odd \cite{QLE}.
\end{widetext}
Next, we recall the expansion of the coth function:
\begin{equation}\label{epsiloncothexpansion}
 \epsilon (\omega, T) = \frac{\hbar \omega}{2} \coth \Bigg(\frac{\hbar \omega}{2 k_B T} \Bigg) = k_B T \Bigg[1 + 2 \sum_{n=1}^\infty \frac{\omega^2}{\omega^2 + \nu_n^2} \Bigg],
\end{equation} where \(\nu_n = \frac{2 \pi n k_B T}{\hbar}\), \(n=1,2,3, \cdots,\) are called the bosonic Matsubara frequencies. Therefore, we may substitute Eq. (\ref{epsiloncothexpansion}) into Eq. (\ref{Edef111112}) and perform the contour integration, to yield the final expression for the average energy. For Drude dissipation, this gives \begin{eqnarray}\label{Eseries}
E(T) = k_B T \Bigg[1+ \sum_{n=1}^{\infty}\frac{2\omega_0^2 +\frac{\nu_n\gamma\omega_{\rm cut}}{\nu_n +\omega_{\rm cut}}}{\nu_n^2 +\omega_0^2 + \frac{\nu_n\gamma \omega_{\rm cut}}{\nu_n +\omega_{\rm cut}}} \Bigg],
\end{eqnarray} where the first term is the classical result and the subsequent terms are quantum corrections. By inspection, the additional terms are all positive definite and therefore, the mean energy of the dissipative quantum oscillator exceeds that of the classical oscillator. A careful reader should note that the series above converges for finite \(\omega_{\rm cut}\), because the large \(n\) terms go as \(\sim 1/n^2\). However, as can be verified easily, in the Ohmic dissipation limit, i.e., \(\omega_{\rm cut} \rightarrow \infty\), the large \(n\) terms go as \(\sim 1/n\) and the series therefore diverges. \\

It is worthwhile to consider the weak-coupling limit, i.e., the limit \(\gamma \rightarrow 0\), which gives
\begin{equation}
E(T)|_{\gamma \rightarrow 0} \approx k_B T \Bigg[1+ \sum_{n=1}^{\infty}\frac{2\omega_0^2}{\nu_n^2 +\omega_0^2} \Bigg].
\end{equation}
This coincides with \(E(T)|_{\gamma \rightarrow 0} \approx \frac{\hbar \omega_0}{2} \coth \big(\frac{\hbar \omega_0}{2 k_B T}\big)\), which is the expected result because in the weak-coupling limit, the dissipative oscillator is equivalent to the quantum harmonic oscillator. It is interesting to note that although we have not put any special limits on the cutoff frequency \(\omega_{\rm cut}\), it has disappeared from the final expression in the weak-coupling limit.

\subsection{Internal energy function}
We now briefly highlight the computation of the internal energy of the dissipative oscillator as was presented earlier in \cite{FordQT0} (see also \cite{bb, FordQT}). The essential logical steps leading to this definition can be understood simply as follows. The idea is to construct the internal energy by subtracting the thermally-averaged energy of the free bath in the absence of the system from that of the full interacting system including the system, bath, and their interactions. For that, one considers the two averages \(U_{\rm Total}(T) = \langle H \rangle\) and \(U_B(t) = \langle H_B \rangle\), where \(H\) and \(H_B\) are given by Eqs. (\ref{Hmodel}) and (\ref{HB}) respectively. The subtlety lies in the fact that both the averages are computed over different distributions. Below, we consider them one by on:
\begin{enumerate}
\item We take a situation where there is no system, just the free bath described by the Hamiltonian \(H_B\) given in Eq. (\ref{HB}). Because the bath is at a thermal equilibrium state for all times, it is described by the equilibrium distribution \(\rho_{H_B}\) given by Eq. (\ref{rhoB}). We then define
\begin{equation}
U_B(T) = \langle H_B \rangle_{\rho_{H_B}} = \frac{{\rm Tr} \big[ H_B e^{-H_B/k_B T}]}{Z_B},
\end{equation} which describes the internal energy of the free bath in its thermal equilibrium state. Since the bath is composed of \(N\) independent quantum oscillators with frequencies \(\{\omega_j\}\), \(j =1,2,\cdots, N\) (see Eq. (\ref{HB})), we have
\begin{equation}
U_B(T) = \sum_{j=1}^N \frac{\hbar \omega_j}{2} \coth \bigg(\frac{\hbar \omega_j}{2 k_B T}\bigg).
\end{equation}

\item Next, consider a parallel situation, where the bath is interacting with the system so that Eq. (\ref{Hmodel}) describes the situation. If the bath has \(N\) degrees of freedom (\(N\) is large), then Eq. (\ref{Hmodel}) describes an \((N+1)\)-oscillator system which, at thermal equilibrium is described by the distribution given in Eq. (\ref{rhoH}). We then define
\begin{equation}
U_{\rm Total}(T) = \langle H \rangle_{\rho_{H}} = \frac{{\rm Tr} \big[ H e^{-H/k_B T}]}{Z},
\end{equation} which is the thermally-averaged energy of the \((N+1)\)-oscillator system. One may introduce a set of transformations (normal-mode coordinates) that map Eq. (\ref{Hmodel}) describing a coupled \((N+1)\)-oscillator system to an uncoupled \((N+1)\)-oscillator system (see \cite{FordQT0} for details). Thereafter, \(U_{\rm Total}(T)\) is given by
\begin{equation}
U_{\rm Total}(T) = \sum_{k=0}^N \frac{\hbar \Omega_k}{2} \coth \bigg(\frac{\hbar \Omega_k}{2 k_B T}\bigg),
\end{equation} where the index \(k\) runs from 0 to \(N\) (as opposed to \(j\) which runs from 1 to \(N\)), and \(\Omega_k\) are the normal-mode frequencies. \\
\end{enumerate}
Now that we have described what the quantities \(U_B(T)\) and \(U_{\rm Total}(T)\) are, we form their difference, i.e.,
\begin{equation}
U(T) = U_{\rm Total}(T) - U_B(T),
\end{equation} which can be interpreted as the internal energy of the dissipative oscillator, in the sense of \cite{FordQT0,FordQT}. The final expression for \(U(T)\) in terms of the susceptibility is
\begin{equation}\label{UT}
U(T) = \frac{1}{\pi} \int_0^\infty \epsilon(\omega, T) {\rm Im} \Bigg[ \frac{d}{d\omega} \ln [\alpha^{(0)}(\omega)] \Bigg]  d\omega,
\end{equation} where \(\epsilon(\omega, T) = \frac{\hbar \omega}{2} \coth \big(\frac{\hbar \omega}{2 k_B T}\big)\). \\
\begin{widetext}
We now perform the integral and express \(U(T)\) as an infinite series. For Drude dissipation, we have \cite{FordQT}
\begin{equation}\label{Drude11111}
 {\rm Im} \Bigg[ \frac{d}{d\omega} \ln [\alpha^{(0)}(\omega)] \Bigg] = \frac{z_+}{z_+^2 + \omega^2} + \frac{z_-}{{z^2_-} + \omega^2} + \frac{\Omega}{\Omega^2 + \omega^2} - \frac{\omega_{\rm cut}}{\omega_{\rm cut}^2 + \omega^2},
\end{equation} where \(\omega_{\rm cut}\) is the Drude cutoff, while the parameters \(z_{\pm}\) are
\begin{equation}\label{zpmdrude}
z_{\pm} = \frac{\Gamma}{2} \pm \sqrt{\frac{\Gamma^2}{4} - \Omega_0^2},
\end{equation} with the parameters \(\Gamma\), \(\Omega_0\), and \(\Omega\) being related to the original parameters of the model, namely \(\gamma\), \(\omega_0\), and \(\omega_{\rm cut}\) as \cite{FordQT}
\begin{equation}\label{parameterDrude}
 \gamma = \frac{ \Gamma[\Omega (\Omega + \Gamma) + \Omega_0^2]}{(\Omega + \Gamma)^2}, \hspace{5mm} \omega_0^2 = \frac{\Omega_0^2 \Omega}{\Omega + \Gamma}, \hspace{5mm} \omega_{\rm cut} = \Omega + \Gamma.
\end{equation}
Let us now substitute Eqs. (\ref{epsiloncothexpansion}) and (\ref{Drude11111}) into Eq. (\ref{UT}) to obtain the internal energy as an infinite series. We perform this integral by closing the contour on the upper half-plane of the complex \(\omega\)-plane. The final result after some simplification reads
\begin{equation}\label{Useries}
U(T) = k_B T \Bigg[1+ \sum_{n=1}^{\infty}\bigg( \frac{\Omega}{\Omega + \nu_n} + \frac{z_+}{z_+ + \nu_n} + \frac{z_-}{z_- + \nu_n} - \frac{\omega_{\rm cut}}{\omega_{\rm cut} + \nu_n} \bigg) \Bigg],
\end{equation}
where we notice that the first term, i.e., \(k_B T\) is just the classical result.\\
\end{widetext}
 Unlike Eq. (\ref{Eseries}), the series given above does not go as \(\sim 1/n^2\) for large \(n\). However, the \(n\)th term of the series for large \(n\) is just
\begin{equation}
u_n \approx \frac{(\Omega + z_+ + z_- - \omega_{\rm cut})}{\nu_n} = \frac{1}{\nu_n} \bigg(-\Gamma + \frac{\Gamma}{2} + \frac{\Gamma}{2}\bigg) = 0,
\end{equation}  thereby ensuring that the series is convergent. In the weak-coupling limit, i.e., with \(\gamma \rightarrow 0\), from Eqs. (\ref{parameterDrude}), we naturally get \(\Gamma \approx 0\), \(\Omega_0 \approx \omega_0\), and \(\Omega \approx \omega_{\rm cut}\). Then, Eq. (\ref{Useries}) becomes
\begin{equation}
U(T)|_{\gamma \rightarrow 0} \approx k_B T \Bigg[1+ \sum_{n=1}^{\infty}\frac{2\omega_0^2}{\nu_n^2 +\omega_0^2} \Bigg].
\end{equation} which gives \(U(T)|_{\gamma \rightarrow 0} \approx \frac{\hbar \omega_0}{2} \coth \Big(\frac{\hbar \omega_0}{2 k_B T}\Big)\), as expected. \\

Interestingly, the thermally-averaged energy of the dissipative oscillator can be obtained in another way, by computing the partition function, based on the Gibbs approach to statistical mechanics. The partition function is obtained via computing Euclidean path integrals, and involves the computation of the influence functional \cite{FV}. The steps are quite mathematically involved, and we refer the reader to \cite{Ingold} for a nice, pedagogical introduction to the subject. The final expression for the partition function reads \cite{Ingold}
\begin{equation}
\mathcal{Z} = \frac{1}{\hbar \omega_0 \beta} \prod_{n=1}^\infty \Bigg[ \frac{\nu_n^2}{\nu^2 + \omega_0^2 + \nu_n \hat{\gamma}(\nu_n)} \Bigg],
\end{equation} where \(\hat{\gamma}(\nu_n) = \gamma \omega_{\rm cut}/(\omega_{\rm cut} + \nu_n)\), for Drude dissipation. \\

\begin{widetext}
From statistical mechanics, the thermally-averaged energy is defined by \(\mathcal{E}(T) = - \frac{\partial}{\partial \beta} \ln \mathcal{Z}\), which reads
\begin{equation}\label{Epath}
\mathcal{E}(T) = k_B T \Bigg[1 + \sum_{n=1}^\infty \frac{2\omega_0^2 + \nu_n \hat{\gamma}(\nu_n)}{\nu_n^2 +\omega_0^2 + \nu_n \hat{\gamma}(\nu_n)} - \sum_{n=1}^\infty \frac{\nu_n^2 \hat{\gamma}'(\nu_n)}{\nu_n^2 +\omega_0^2 + \nu_n \hat{\gamma}(\nu_n)} \Bigg].
\end{equation}
\end{widetext}
Obviously, Eqs. (\ref{Eseries}) and (\ref{Epath}) do not match, except in the special case where \(\hat{\gamma}(\nu_n)\) is a constant, as is for Ohmic dissipation. A careful inspection tells us that \(\mathcal{E}(T) - E(T) > 0\), i.e., the average energy obtained from the partition function method exceeds that obtained from the fluctuation-dissipation theorem. However, when the cutoff frequency \(\omega_{\rm cut}\) is much larger than all the relevant energy scales of the system, i.e., \(\omega_{\rm cut} >> \omega_0, \gamma\), the second summation in Eq. (\ref{Epath}) becomes small and in that case, \(\mathcal{E}(T) \approx E(T)\). One may also verify that in the weak-coupling limit, i.e., for \(\gamma \rightarrow 0\), \(\mathcal{E}(T)\) gives the expected expression, i.e., Eq. (\ref{11111}).\\

\begin{figure}
\begin{center}
\includegraphics[scale=0.80]{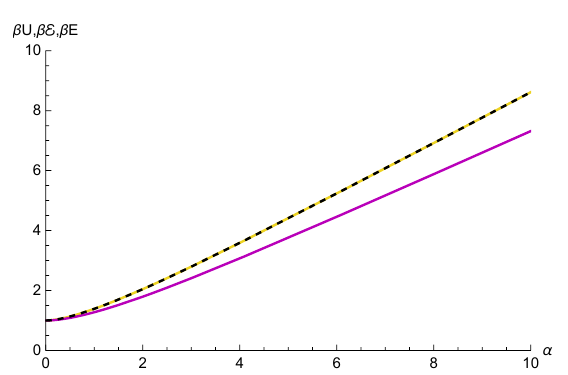}
\caption{Dimensionless energy functions \(\beta U\) (black-dashed), \(\beta \mathcal{E}\) (yellow-solid), and \(\beta E\) (violet-solid) for the dissipative oscillator, as a function \(\alpha = \beta \hbar \omega_0\), for \(\gamma = \omega_0\) and \(\omega_{\rm cut} = 10 \omega_0\). We have taken \(\omega_0 = 1\) and have considered the first 10,000 terms of Eqs. (\ref{Eseries}), (\ref{Useries}), and (\ref{Epath}).}
\label{EcalEcomp}
\end{center}
\end{figure}

\begin{figure}
\begin{center}
\includegraphics[scale=0.80]{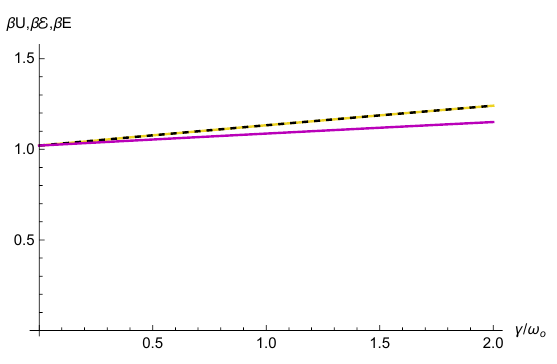}
\caption{Dimensionless energy functions \(\beta U\) (black-dashed), \(\beta \mathcal{E}\) (yellow-solid), and \(\beta E\) (violet-solid) for the dissipative oscillator, as a function \(\gamma/\omega_0\), for \(\alpha = \beta \hbar \omega_0 = 0.5\) and \(\omega_{\rm cut} = 10 \omega_0\). We have taken \(\omega_0 = 1\) and have considered the first 10,000 terms of Eqs. (\ref{Eseries}), (\ref{Useries}), and (\ref{Epath}).}
\label{EcalEcomp1}
\end{center}
\end{figure}

 In Figs. (\ref{EcalEcomp}) and (\ref{EcalEcomp1}), we have plotted all the energy functions in units of \(k_B T\). Strikingly, the energy functions \(U\) and \(\mathcal{E}\) are equivalent, although they are obtained using different methods. This shall be called the internal energy function, while the function \(E\) obtained in the previous subsection will be called the mean energy function. It is observed that \(U, \mathcal{E} > E\) can also be verified by inspection of their expressions in the form of infinite series. All the energy functions show similar qualitative trends, i.e., they can be enhanced by increasing the strength of system-bath coupling or by increasing the ratio \(\alpha = \hbar \omega_0/k_B T\) (notation not to be confused with susceptibility), the latter implying that one is moving deeper into the quantum regime. Conversely, one can see from Fig. (\ref{EcalEcomp}) that for \(k_B T >> \hbar \omega_0\), i.e., \(\alpha \rightarrow 0\), all the energy functions return to the usual classical value \(k_B T\), as is expected. Moreover, from Fig. (\ref{EcalEcomp1}), one observes that as we take \(\gamma \rightarrow 0\), the energy functions saturate to the corresponding value obtained from Eq. (\ref{11111}) in the weak-coupling limit. Thus, the energy functions of the dissipative oscillator always exceed the energy of the oscillator as obtained in the weak-coupling assumption, given by Eq. (\ref{11111}). In this sense, Eqs. (\ref{Eseries}) and (\ref{Useries}), the latter being equal to (\ref{Epath}) are appropriate generalizations of the energy of the harmonic oscillator, while it is coupled to a heat bath with arbitrary coupling strength. It is noteworthy that for Ohmic dissipation, i.e., for \(\omega_{\rm cut} \rightarrow \infty\), \(E = U = \mathcal{E}\), i.e., the energy functions are all equivalent but diverge as \(1/n\) for large \(n\). \\

 We present the numerical values of the three energy functions for varying \(\alpha\) in Table (I). It can be observed that the functions \(U(T)\) and \(\mathcal{E}(T)\) agree, up to multiple decimal places. However, the mean energy function, i.e., \(E(T)\), assumes values that are notably distinct from the other two energy functions. We have taken \(\gamma = \omega_0\), \(\omega_{\rm cut} = 10 \omega_0\), and \(\omega_0 = 1\).

\begin{table*}[t]
\caption{Numerical values of the energy functions for the dissipative oscillator}
\centering
\begin{tabular}{| c | c | c | c |}
\hline
$ \alpha = \hbar \omega_0/k_B T$ & $ E$ &  $ U $ & $ \mathcal{E} $   \\ \hline
0.1   &   1.00457  &  1.00793      &  1.00793   \\

0.3     &       1.03540   &   1.05693      &  1.05693  \\

0.5      &      1.08702  &    1.13288   &  1.13288  \\

1.5      &       1.51726      & 1.69975    & 1.69976  \\

2.5      &      2.09352      &  2.41169    &  2.41171  \\

3.5      &      2.73638      &  3.18759   &  3.18763  \\

5.5      &       4.10794     &  4.82152   &  4.82158  \\

7.5      &      5.52459     &  6.49860   &  6.49869 \\ \hline
\end{tabular}
\end{table*}

\subsection{Quantum counterpart of energy equipartition theorem}
Let us briefly discuss the quantum counterpart of energy equipartition theorem in the present context, mainly revisiting the material presented in earlier works  \cite{jarzy1,jarzy2,jarzy3,jarzy4,jarzy45,jarzy5,jarzy6,kaur,kaur1,kaur2,kaur3}. It may straightforwardly be shown that \cite{kaur}
\begin{equation}\label{P2generic}
P_E(\omega) = \frac{m}{\pi} {\rm Im} [\alpha^{(0)}(\omega + i 0^+)] \bigg(1 + \frac{\omega_0^2}{\omega^2}\bigg) \omega,
\end{equation}
is both positive definite and normalized, meaning that it is a genuine probability distribution function [see App. (\ref{appendixA})]. Thus, we may rewrite Eq. (\ref{Edef111}) as
\begin{equation}\label{Edef11111}
E(T) = \int_0^\infty \epsilon(\omega,T) P_E (\omega) d\omega.
\end{equation}
Eq. (\ref{Edef111}) can be interpreted along the lines of the so-called quantum counterpart of energy equipartition theorem: the quantity \(\epsilon(\omega,T) P_E(\omega) d\omega\) corresponds to the contribution to the mean energy \((E(T))\) of the dissipative oscillator coming from bath degrees of freedom lying in the frequency window between \(\omega\) and \(\omega + d\omega\). As can be readily verified using Eqs. (\ref{alphadef11}) and (\ref{P2generic}), putting \(\tilde{\mu}(\omega) \rightarrow 0\), i.e., \(\gamma \rightarrow 0\), implies \(P_E(\omega) |_{\gamma \rightarrow 0} =  \approx \delta(\omega - \omega_0) + \delta(\omega + \omega_0).\). Substituting this into Eq. (\ref{Edef11111}) simply gives \(E(T)|_{\gamma \rightarrow 0} \approx \frac{\hbar \omega_0}{2} \coth \big(\frac{\hbar \omega_0}{2 k_B T}\big)\), which is the expected answer [Eq. (\ref{11111})]. One may in this case, split the mean energy into the kinetic and potential energy parts, and obtain the quantum counterpart of energy equipartition theorem individually for the kinetic and potential energies as has been studied in \cite{jarzy3,kaur}.\\

We now consider the internal energy function (\(U(T)\)). In the light of the quantum counterpart of energy equipartition theorem, Eq. (\ref{UT}) acquires a special interpretation. As has been recently demonstrated \cite{kaur2}, the quantity \(\frac{1}{\pi} {\rm Im} \Big[ \frac{d}{d\omega} \ln [\alpha^{(0)}(\omega)] \Big] \) has all the properties of a probability distribution function [see App. (\ref{appendixB})] call it \(P_U(\omega)\). Therefore, Eq. (\ref{UT}) can be rewritten as
\begin{equation}\label{Uqc}
U(T) = \int_0^\infty \epsilon(\omega,T) P_U(\omega) d\omega,
\end{equation}
which means \(\epsilon(\omega,T) P_U(\omega) d\omega\) corresponds to the contribution to the internal energy of the dissipative oscillator coming from bath degrees of freedom lying in the frequency window between \(\omega\) and \(\omega + d\omega\). In the classical limit, \(\epsilon(\omega,T) \rightarrow k_B T\) and because \(P_U(\omega)\) is normalized due to it being a probability distribution function, we have \(U(T)|_{k_B T >> \hbar \omega_0} \approx k_B T\). We emphasize to the reader that the quantities \(E(T)\) and \(U(T)\) are very different, and thus, \(P_E(\omega)\) and \(P_U(\omega)\) are different probability distribution functions. It is both surprising and fascinating that both \(E(T)\) and \(U(T)\) can be interpreted along the lines of the quantum counterpart of energy equipartition theorem \footnote{We should further clarify that in \cite{jarzy1,jarzy2,jarzy3,jarzy4,jarzy45,jarzy5,jarzy6,kaur,kaur1,kaur3}, the quantum counterpart of energy equipartition theorem has been explicitly demonstrated for \(E(T)\). However, in \cite{kaur2}, the quantum counterpart of energy equipartition theorem was shown for \(U(T)\), although there it was referred to as mean energy or average energy, because precise difference between \(U(T)\) and \(E(T)\) was noticed by us later.}, and lead to the correct classical-limit result, i.e., \(k_B T\) for \(\hbar \rightarrow 0\). Furthermore, in the weak-coupling limit, one can easily verify that \(P_U(\omega) |_{\gamma \rightarrow 0}   \approx \delta(\omega - \omega_0) + \delta(\omega + \omega_0)\), meaning that \(U(T)|_{\gamma \rightarrow 0} = \frac{\hbar \omega_0}{2} \coth \big(\frac{\hbar \omega_0}{2 k_B T}\big)\). \\

\begin{figure}
\begin{center}
\includegraphics[scale=0.80]{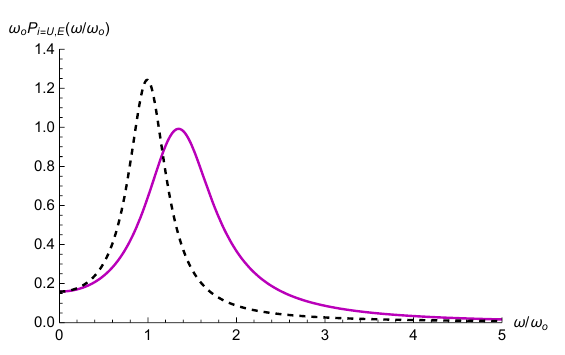}
\caption{The distribution functions \(\omega_0 P_U(\omega/\omega_0)\) (black-dashed) and \(\omega_0 P_E(\omega/\omega_0)\) (violet-solid), as a function of the rescaled heat-bath frequency \(\omega/\omega_0\), for \(\gamma = 0.5 \omega_0\) and \(\omega_{\rm cut} = 5 \omega_0\). We have taken \(\omega_0 = 1\).}
\label{tilde rho vs tilde omega}
\end{center}
\end{figure}

\begin{widetext}

We also note that \(E(T) = U(T)\) for Ohmic dissipation, for which some simple algebra reveals that
\begin{eqnarray}
{\rm Im}\Bigg[\frac{d}{d\omega}\ln[\alpha^{(0)}(\omega)]\Bigg] &=&   {\rm Im} [\alpha^{(0)}(\omega)] \bigg(1 + \frac{\omega_0^2}{\omega^2}\bigg) m \omega = \frac{\gamma (\omega^2 + \omega_0^2)}{(\omega^2 - \omega_0^2)^2 + (\gamma \omega)^2},
\end{eqnarray} therefore giving \(P_E(\omega) = P_U(\omega)\). This appears to arise due to the oversimplified form of \(\tilde{\mu}(\omega)\) for Ohmic dissipation. In that case, it is a real constant, i.e., it lacks an imaginary part, and its derivative is zero. In general however, the result stated above is not true. For instance, in case of Drude dissipation, where \(\tilde{\mu} (\omega)\) is a complex-valued function of \(\omega\), the integrands of Eqs. (\ref{UT}) and (\ref{Edef111}) simply do not match. It is imperative to compare the probability distributions \(P_U(\omega)\) and \(P_E(\omega)\). In Fig. (\ref{tilde rho vs tilde omega}), we have plotted both \(P_E(\omega)\) and \(P_U(\omega)\), in dimensionless form, as a function of the rescaled heat-bath frequency \(\omega/\omega_0\), for Drude dissipation, and for the same values of parameters. Both of them admit a single peak, which is just the most probable frequency for each case. As it turns out, the most probable frequencies for \(E(T)\) and \(U(T)\) are different.

\section{Dissipative magneto-oscillator}\label{magSec}
Let us now consider a three-dimensional generalization of the dissipative oscillator, in the presence of an applied magnetic field. This is the so-called dissipative magneto-oscillator and is described by the following Hamiltonian:
\begin{eqnarray}\label{H}
   H =&& \frac{(\mathbf{p} - \frac{e}{c} \mathbf{A})^2}{2m} + \frac{m \omega_0^2 \mathbf{r}^2}{2}  + \sum_{j=1}^N\bigg[\frac{\mathbf{p}_j^2}{2m_j} + \frac{1}{2}m_j \omega_j^2 \big( \mathbf{q}_j -\mathbf{r} \big)^2 \bigg],
\end{eqnarray}
where $\mathbf{p}$ and $\mathbf{r}$ are the momentum and position operators of the system, $\mathbf{p}_j$ and $\mathbf{q}_j$ are the corresponding variables for the \(j\)-th  oscillator of the heat bath, and $\mathbf{A}$ is the vector potential. The usual commutation relations between coordinates and momenta hold. Integrating out the reservoir variables from the Heisenberg equations of motion and assuming that the system and the bath were at thermal equilibrium initially, one obtains the quantum Langevin equation:
\begin{equation}\label{eqnm}
  m \ddot{\mathbf{r}}(t) + \int_{-\infty}^{t} \mu(t - t') \dot{\mathbf{r}}(t') dt' + m \omega_0^2 \mathbf{r}(t)-\frac{e}{c}(\dot{\mathbf{r}}(t) \times \mathbf{B}) = \mathbf{F}(t),
\end{equation}
where \(\mu(t)\) is the dissipation kernel defined in Eq. (\ref{1}), and \(\mathbf{F}(t)\) is the random force with expression obtained by generalizing Eq. (\ref{noise}) to three dimensions.
The noise in a certain spatial direction is uncorrelated (and commutes) with the noise in any other direction. In each spatial direction, the noise at time \(t\) and that at \(t'\) are correlated as in Eq. (\ref{symmetricnoisecorrelation1}), while the commutator is given in (\ref{noisecommutator1}). \\

As with the one-dimensional case, we can solve the quantum Langevin equation by adding a perturbing force \(\mathbf{f}(t)\) and considering its Fourier transform:
\begin{equation}
\tilde{r}_{\rho}(\omega)=\alpha_{\rho\sigma}(\omega)[\tilde{f}_{\sigma}(\omega)+\tilde{F}_{\sigma}(\omega)],
\end{equation}
where
\begin{equation}
  \alpha_{\rho\sigma}(\omega)=\frac{[\lambda(\omega)]^2\delta_{\rho\sigma}-(\omega e/c)^2B_{\rho}B_{\sigma}-i\frac{e\omega}{c}\lambda(\omega)\epsilon_{\rho\sigma\eta}B_{\eta}}{\det[D(\omega)]},
\end{equation} with
\begin{eqnarray}
&&\det[D(\omega)]=\lambda(\omega)\Big[\lambda(\omega)^2-(e\omega/c)^2B^2\Big], \\
&&\lambda(\omega)=\Big[m(\omega_0^2-\omega^2)-i\omega\tilde{\mu}(\omega)\Big],\\
&&\tilde{r}_{\rho}(\omega)=\int_{-\infty}^{\infty}dt e^{i\omega t} r_{\rho}(t), \\
&&\tilde{\mu}(\omega)=\int_{0}^{\infty}dt e^{i\omega t} \mu(t),
\end{eqnarray}
and $\epsilon_{\rho\sigma\eta}$ is the Levi-Civita antisymmetric tensor. The Greek indices stand for three spatial directions (i.e., $\rho , \sigma, \eta = x,y,z$) and the Einstein summation convention is used. The generalized susceptibility tensor $\alpha_{\rho \sigma}(\omega)$ determines the dynamics of the system.

\subsection{Mean energy function}
One may now compute the mean energy function, as well as the internal energy function of the dissipative magneto-oscillator. For mean energy, we need to compute equal-time correlation functions \(\langle \dot{x}^2 \rangle\), \(\langle \dot{y}^2 \rangle\), \(\langle \dot{z}^2 \rangle\), \(\langle x^2 \rangle\), \(\langle y^2 \rangle\), and \(\langle z^2 \rangle\). They can be found by invoking a three-dimensional generalization of the Callen-Welton fluctuation-dissipation theorem \cite{callenwelton,Kubo,case,bb}:

\begin{eqnarray}
&& \frac{1}{2}\langle r_{\rho}(t) r_{\sigma}(t') + r_{\sigma}(t') r_{\rho}(t)\rangle \label{lkjhg}  \\
  &=& \frac{\hbar}{\pi} \int_{0}^{\infty} {\rm Im} [\alpha^s_{\rho\sigma}(\omega)]\coth\bigg(\frac{\hbar \omega}{2k_BT}\bigg) \cos[\omega(t-t')] d\omega -\frac{\hbar}{\pi} \int_{0}^{\infty} {\rm Re} [\alpha^a_{\rho\sigma}(\omega)]\coth\bigg(\frac{\hbar \omega}{2k_BT}\bigg) \sin[\omega(t-t')] d\omega \nonumber
\end{eqnarray}
 where $\alpha_{\rho \sigma}^s (\omega)$ and $\alpha_{\rho \sigma}^a (\omega)$ are the symmetric and antisymmetric parts of the generalized susceptibility tensor. For
definiteness, we now choose the direction of the magnetic field as the \(z\)-direction in calculations throughout this paper. Due to the cylindrical symmetry of the system, the only non-zero elements of the generalized susceptibility tensor $\alpha_{\rho \sigma}^s(\omega)$ are $\alpha_{xx}^s(\omega)$, $\alpha_{yy}^s(\omega)$, $\alpha_{zz}^s(\omega)$, $\alpha_{xy}(\omega)$, and $\alpha_{yx}(\omega)$. The non-vanishing elements are
\begin{eqnarray}
\alpha_{xx}^s(\omega)=\alpha_{yy}^s(\omega)=\frac{[\lambda(\omega)]^2}{\det D(\omega)}, \hspace{10mm} \alpha_{zz}^s(\omega)=\frac{1}{\lambda(\omega)}, \hspace{10mm} \alpha_{xy}(\omega)=-\alpha_{yx}(\omega)= -i\omega\frac{e}{c}\frac{B\lambda(\omega)}{\det D(\omega)}.
\end{eqnarray}
Thus, we have the following equal-time position autocorrelation functions:
\begin{eqnarray}\label{x}
\langle x^2 \rangle= \langle y^2 \rangle = \frac{\hbar}{\pi}\int_{0}^{\infty}d\omega {\rm Im}[\alpha_{xx}^s(\omega)]\coth\Big(\frac{\hbar\omega}{2k_BT}\Big), \hspace{7mm} \langle z^2 \rangle =\frac{\hbar}{\pi}\int_{0}^{\infty}d\omega {\rm Im}[\alpha_{zz}^s(\omega)]\coth\Big(\frac{\hbar\omega}{2k_BT}\Big).
\end{eqnarray}
Next, in order to obtain the velocity correlation functions, we differentiate Eq. (\ref{lkjhg}) first with respect to \(t\), then with respect to \(t'\), before finally setting \(t = t'\). This gives
\begin{equation}\label{y}
  \langle \dot{{x}}^2 \rangle=\langle \dot{{y}}^2 \rangle = \frac{\hbar}{\pi} \int_{0}^{\infty} d\omega \omega^2  {\rm Im}[\alpha^s_{xx}(\omega)]\coth\bigg(\frac{\hbar \omega}{2k_BT}\bigg), \hspace{7mm} \langle \dot{{z}}^2 \rangle= \frac{\hbar}{\pi} \int_{0}^{\infty} d\omega \omega^2  {\rm Im}[\alpha^s_{zz}(\omega)]\coth\bigg(\frac{\hbar \omega}{2k_BT}\bigg).
\end{equation}
\end{widetext}
Using these equal-time correlation functions, we may now compute the mean energy of the system, which is the average of the system Hamiltonian \(H_S\) performed over the noise ensemble. The final result for the mean energy reads
\begin{equation}\label{E3dp}
E(T) = \int_0^\infty  \epsilon^{(3)}(\omega,T) P_E (\omega) d\omega,
\end{equation} where \(\epsilon^{(3)}(\omega,T) = \frac{3 \hbar \omega}{2} \coth \big(\frac{\hbar \omega}{2 k_B T}\big)\) is the mean energy of a three-dimensional oscillator of the heat bath. The function \(P_E(\omega)\) is found to be
\begin{equation}
P_E (\omega) = \frac{m \omega}{3 \pi} {\rm Im} [\alpha^s_{\sigma \sigma} (\omega)] \bigg(1 + \frac{\omega_0^2}{\omega^2}\bigg),
\end{equation} where \(\alpha^s_{\sigma \sigma} (\omega)\) is the symmetric part of the susceptibility tensor. In \cite{kaur}, it was demonstrated that \(P_E(\omega)\) has all the properties of a probability distribution function on the real half-line, i.e., \(\omega \in [0,\infty)\). This allows one to interpret Eq. (\ref{E3dp}) along the lines of the quantum counterpart of energy equipartition theorem. In Figs. (\ref{fig:test1}), (\ref{fig:test2}), and (\ref{fig:test3}), we have plotted the probability distribution function \(P_E(\omega)\). \\

Now, the exact expression for energy can be found by computing the integral on the frequency plane. For Drude dissipation, the final expression reads (see also \cite{kaur})
\begin{widetext}
\begin{equation}\label{E3dSeries}
E(T) = k_B T \Bigg[3  + 2 \sum_{n=1}^{\infty}\frac{ \big( \nu_n^2 +\omega_0^2 + \frac{\nu_n\gamma \omega_{\rm cut}}{\nu_n +\omega_{\rm cut}}\big)\times \big( 2 \omega_0^2 +(\omega_c\nu_n)^2 + \frac{\nu_n\gamma\omega_{\rm cut}}{\nu_n +\omega_{\rm cut}}\big)}{ \big( \nu_n^2 +\omega_0^2 + \frac{\nu_n\gamma \omega_{\rm cut}}{\nu_n +\omega_{\rm cut}}\big)^2+(\omega_c\nu_n)^2} + \sum_{n=1}^\infty \frac{2 \omega_0^2 + \frac{\nu_n\gamma\omega_{\rm cut}}{\nu_n +\omega_{\rm cut}}}{ \nu_n^2 +\omega_0^2 + \frac{\nu_n\gamma \omega_{\rm cut}}{\nu_n +\omega_{\rm cut}}} \Bigg],
\end{equation}
where the first term, i.e., \(3 k_B T\) is the classical contribution. The remaining terms are positive definite, indicating that the mean energy of the quantum dissipative magneto-oscillator exceeds that of its classical counterpart. Clearly, the mean energy can be influenced by the magnetic field (via cyclotron frequency \(\omega_c\)), dissipation strength \(\gamma\), harmonic trap frequency \(\omega_0\), and cutoff frequency \(\omega_{\rm cut}\). It can be verified that the terms of the series above go as \(\sim 1/n^2\) for large \(n\), but for \(\omega_{\rm cut} \rightarrow \infty\), the the series diverges as the terms go as \(\sim 1/n\) for large \(n\). We shall later compare the values of the mean energy function as obtained from Eq. (\ref{E3dSeries}) with the internal energy function. \\
\end{widetext}

\subsection{Internal energy function}
Now, let us consider the internal energy function which was studied for the one-dimensional case earlier. In the present case, it is defined in the same way as before: it is the difference between the thermally-averaged value of the total Hamiltonian of the system + bath setup, in the canonical equilibrium state of the bath, and the thermally-averaged value of the bath Hamiltonian, in the canonical equilibrium state of the bath (in the absence of the system), i.e., \(U(T) := U_{\rm Total}(T) - U_B(T)\).

\begin{widetext}
We refer the reader to \cite{molu} and references therein, for the detailed calculation and quote the final expression below:
\begin{eqnarray}
  U(T) &=&\frac{1}{\pi}\int_{0}^{\infty}d\omega \epsilon^{(3)}(\omega,T){\rm Im}\Big[\frac{d}{d\omega}\ln[\det\alpha(\omega)]\Big] \nonumber \\
  &=& U_0(T) + \Delta U (T),
\end{eqnarray} where the first term is simply three times of Eq. (\ref{UT}) and the second term above carries the contribution due to the magnetic field, which reads
\begin{equation}\label{DeltaF}
  \Delta U (T) = - \frac{1}{3\pi} \int_{0}^{\infty} \epsilon^{(3)}(\omega,T) {\rm Im} \Bigg[ \frac{d}{d\omega} \ln \bigg(1 - \bigg(\frac{eB \omega \alpha^{(0)}(\omega)}{c}\bigg)^2 \bigg) \Bigg] d\omega.
\end{equation}
 Here, \(\alpha^{(0)}(\omega)\) is the susceptibility defined in Eq. (\ref{alphadef11}). Using this, one gets the following expression for the internal energy function:
 \begin{equation}\label{U3dqc}
 U(T) = \int_0^\infty \epsilon^{(3)}(\omega,T) P_U(\omega) d\omega,
 \end{equation}
 where \(\epsilon^{(3)}(\omega,T)\) is the thermally-averaged energy of a three-dimensional quantum oscillator, and
 \begin{eqnarray}
 P_U(\omega) = \frac{1}{\pi} \Bigg[-\frac{\omega_{\rm cut}}{\omega^2+\omega_{\rm cut}^2} + \frac{\Omega}{\omega^2+{\Omega}^2} + \frac{1}{3}\bigg( \frac{z_+}{\omega^2+ {z_+}^2}+\frac{{z_-}}{\omega^2+{z_-}^2} + \frac{\Omega_1}{\omega^2+\Omega_1^2} +\frac{\Omega_1^*}{\omega^2+{\Omega_1^*}^2}+\frac{\Omega_2}{\omega^2+{\Omega_2}^2}+\frac{\Omega_2^*}{\omega^2+{\Omega_2^*}^2}\bigg)\Bigg]. \nonumber \\ \label{Pthreedim}
\end{eqnarray}
Here \(z_{\pm}\) are defined using Eqs. (\ref{zpmdrude}), while we have
\begin{eqnarray}
  \Omega_1 = \Big\lbrack\frac{\Gamma}{2}+\Big(\frac{b-a}{2}\Big)^{\frac{1}{2}}\Big\rbrack-i\Big\lbrack\frac{\omega_c}{2}+\Big(\frac{b+a}{2}\Big)^{\frac{1}{2}}\Big\rbrack, \hspace{5mm}  \Omega_2=\Big\lbrack\frac{\Gamma}{2}-\Big(\frac{b-a}{2}\Big)^{\frac{1}{2}}\Big\rbrack-i\Big\lbrack\frac{\omega_c}{2}-\Big(\frac{b+a}{2}\Big)^{\frac{1}{2}}\Big\rbrack,
\end{eqnarray}
  with
  \begin{equation}
    a =\Big(\frac{\omega_c}{2}\Big)^2+\Big(\Omega_0^2-\frac{\Gamma^2}{4}\Big), \hspace{3mm} {\rm and}  \hspace{3mm} b=\Big\lbrack a^2+\Big(\frac{\Gamma\omega_c}{2}\Big)^2\Big\rbrack^{\frac{1}{2}}.
  \end{equation}
 Interestingly, one can interpret Eq. (\ref{U3dqc}) along the lines of the quantum counterpart of energy equipartition theorem, because the function \(P_U(\omega)\) is positive definite (see \cite{kaur2}) and can also be verified to be normalized by performing an explicit integration. Thus, \(\epsilon^{(3)}(\omega,T)P_U(\omega) d\omega\) is the contribution to the internal energy of the system coming from bath oscillators in the frequency range between \(\omega\) and \(\omega + d\omega\). \\

\begin{figure}
\subfloat[\(\omega_c = 0.5 \omega_0\)\label{fig:test1}]
  {\includegraphics[width=0.4\linewidth]{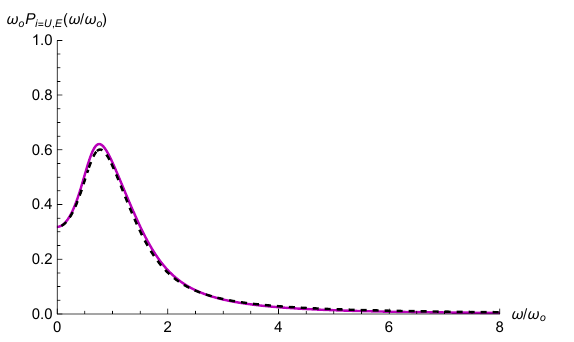}}\hfill
\subfloat[\(\omega_c = 2.5 \omega_0\).\label{fig:test2}]
  {\includegraphics[width=0.4\linewidth]{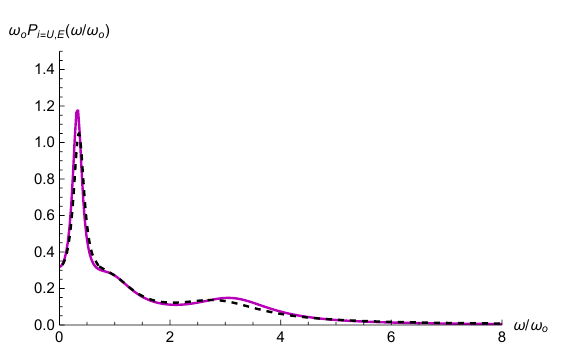}}\hfill
\subfloat[\(\omega_c = 5.0 \omega_0\).\label{fig:test3}]
  {\includegraphics[width=0.4\linewidth]{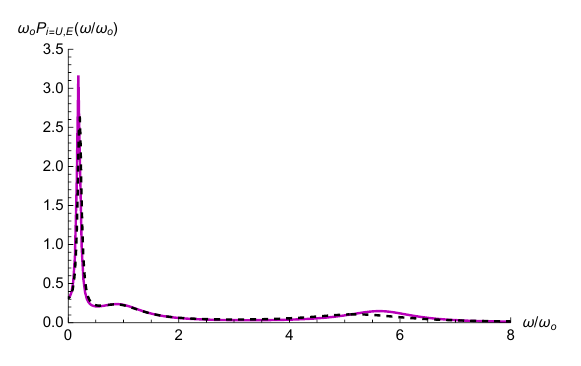}}
\caption{The distribution functions \(\omega_0 P_U(\omega/\omega_0)\) (black-dashed) and \(\omega_0 P_E(\omega/\omega_0)\) (violet-solid), as a function of the rescaled heat-bath frequency \(\omega/\omega_0\), for \(\gamma = 0.5 \omega_0\), \(\omega_{\rm cut} = 10 \omega_0\), and various values of cyclotron frequency. We have taken \(\omega_0 = 1\).}
\end{figure}

In Figs. (\ref{fig:test1}), (\ref{fig:test2}), and (\ref{fig:test3}), we have plotted the probability distribution functions \(P_U(\omega)\) and \(P_E(\omega)\) for different values of the magnetic field strength. Both the probability distribution functions exhibit similar behavior, meaning that the mean energy and internal energy functions behave similarly as far as the quantum counterpart of the energy equipartition theorem is concerned. However, the functions do not exactly coincide, indicating the fact that the energy functions \(U(T)\) and \(E(T)\) are distinct and inequivalent unless, of course, one opts for Ohmic dissipation, for which they are equivalent as in the one-dimensional case discussed earlier. However, one can observe that for small values of the magnetic field, the probability distribution functions have just one peak, indicating the existence of a most probable frequency (distinct for \(U(T)\) and \(E(T)\)). However, upon increasing the magnetic field, one notices the appearance of up to three most probable frequencies for each energy function. It should be pointed out that the appearance of the multiple peaks (most probable frequencies) in the context of the mean energy for an oscillator in a magnetic field was studied earlier in \cite{kaur,kaur3}. \\

Having discussed the quantum counterpart of energy equipartition theorem, let us now compute the internal energy function exactly, by performing the contour integration of Eq. (\ref{U3dqc}) explicitly. The final answer reads
\begin{equation}\label{Useries3d}
U(T) = 3 k_B T \Bigg[1+ \sum_{n=1}^{\infty}\bigg( \frac{\Omega}{\Omega + \nu_n} - \frac{\omega_{\rm cut}}{\omega_{\rm cut} + \nu_n} \bigg) + \frac{1}{3} \sum_{n=1}^\infty \bigg( \frac{z_+}{z_+ + \nu_n} + \frac{z_-}{z_- + \nu_n} + \frac{\Omega_1}{\Omega_1 + \nu_n} + \frac{\Omega_1^*}{\Omega_1^* + \nu_n} + \frac{\Omega_2}{\Omega_2 + \nu_n} + \frac{\Omega_2^*}{\Omega_2^* + \nu_n} \bigg) \Bigg],
\end{equation} where \(\nu_n\) are the bosonic Matsubara frequencies, as before. In Fig. (\ref{3Dalpha}), we have plotted the internal energy function \(U(T)\), together with the mean energy function \(E(T)\), in units of \(k_BT\), as a function of the ratio \(\alpha = \hbar \omega_0/k_B T\). We have also plotted the energy function \(\mathcal{E}(T)\) obtained from the partition for comparison, whose expression reads
\begin{equation}\label{Epath3d}
\mathcal{E}(T) = \mathcal{E}_{xy}(T) + \mathcal{E}_z(T),
\end{equation} where
\begin{equation}
 \mathcal{E}_{xy}(T) = k_B T \Bigg[2 -\sum _{n=1}^{\infty} \frac{\nu_{n} ^5 \bigg(\frac{2  \omega_{c}^2 \nu_{n} +2 \left(\frac{\gamma  \omega_{\rm cut}}{\omega_{\rm cut}+\nu_{n} }-\frac{\gamma  \omega_{\rm cut} \nu_{n} }{(\omega_{\rm cut}+\nu_{n} )^2}+2 \nu_{n} \right) \left(\frac{\gamma  \omega_{\rm cut} \nu_{n} }{\omega_{\rm cut}+\nu_{n} }+\nu_{n} ^2+  \omega_0^2\right)}{\nu_{n} ^4}-\frac{4 \left(  \omega_{c}^2 \nu_{n} ^2+\left(\frac{\gamma  \omega_{\rm cut} \nu_{n} }{\omega_{\rm cut}+\nu_{n} }+\nu_{n} ^2+  \omega_0^2\right)^2\right)}{\nu_{n} ^5}\bigg)}{( \omega_{c} \nu_{n} )^2+\left(\frac{\nu_{n}  (\gamma  \omega_{\rm cut})}{\omega_{\rm cut}+\nu_{n} }+\nu_{n} ^2+ \omega_0^2\right)^2}\Bigg], \nonumber
\end{equation}
\begin{equation}
 \mathcal{E}_{z}(T) = k_B T \Bigg[1 + \sum_{n=1}^\infty \frac{2\omega_0^2 + \frac{\nu_n \gamma \omega_{\rm cut}}{\omega_{\rm cut} + \nu_n}
}{\nu_n^2 +\omega_0^2 + \frac{\nu_n \gamma \omega_{\rm cut}}{\omega_{\rm cut} + \nu_n}
} + \sum_{n=1}^\infty \frac{ \frac{\nu_n^2 \gamma \omega_{\rm cut}}{(\omega_{\rm cut} + \nu_n)^2}
}{\nu_n^2 +\omega_0^2 + \frac{\nu_n\gamma \omega_{\rm cut}}{\nu_n +\omega_{\rm cut}}} \Bigg]. \nonumber
\end{equation}
It is obtained by adding Eq. (\ref{Epath}) to the thermally-averaged energy for the two-dimensional magneto-oscillator determined from the partition function presented in \cite{SDGPath}. Unlike the one-dimensional case, one can observe that \(\mathcal{E}(T)\) differs slightly from the internal energy function \(U(T)\). In Fig. (\ref{3Dgamma}), we have plotted all the energy functions as a function of \(\gamma\) and the qualitative behavior observed matches that of the one-dimensional dissipative oscillator. It is interesting to note that in the case of the magneto-oscillator, the energy function \(\mathcal{E}(T)\), which is obtained from the partition function does not coincide with the internal energy function exactly, unlike the one-dimensional case. This is an effect due to the applied magnetic field, because if there was no magnetic field, the three-dimensional case would be merely just three copies of the one-dimensional dissipative oscillator, and then the functions \(\mathcal{E}(T)\) and \(U(T)\) are expected to coincide, as can be seen from Table (I). In Table (II), we present the numerical values of all three energy functions for a few values of \(\alpha\) for comparison. We have taken \(\gamma = \omega_0\), \(\omega_c = 2.5 \omega_0\), \(\omega_{\rm cut} = 10 \omega_0\), and \(\omega_0 = 1\).\\
\end{widetext}

\begin{table*}[t]
\caption{Numerical values of the energy functions for the dissipative magneto-oscillator}
\centering
\begin{tabular}{| c | c | c | c |}
\hline
$ \alpha = \hbar \omega_0/k_B T$ & $ E$ &  $ U $ & $ \mathcal{E} $   \\ \hline
0.1   &   3.01880  &  3.02870      &  3.02930   \\

0.3     &       3.15098   &     3.21200    &  3.21470  \\

0.5      &      3.37795  &    3.50310   &  3.51420  \\

1.5      &       5.23727      & 5.68050    & 5.76270  \\

2.5      &      7.54411      &  8.29530    &  8.44880  \\

3.5      &      9.98530      &  11.0469   &  11.26310 \\

5.5      &       15.02940     &  16.71580   &  17.04670  \\

7.5      &      20.17460     &  22.48770  &  22.92800 \\ \hline
\end{tabular}
\end{table*}

\begin{figure}
\begin{center}
\includegraphics[scale=0.80]{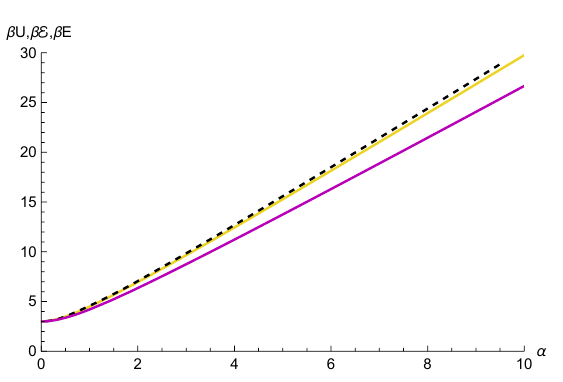}
\caption{Dimensionless energy functions \(\beta U\) (black-dashed), \(\beta \mathcal{E}\) (yellow-solid), and \(\beta E\) (violet-solid) for the dissipative magneto-oscillator, as a function \(\alpha = \beta \hbar \omega_0\), for \(\gamma = \omega_0\), \(\omega_c = 2.5 \omega_0\), and \(\omega_{\rm cut} = 10 \omega_0\). We have taken \(\omega_0 = 1\) and have considered the first 10,000 terms of Eqs. (\ref{E3dSeries}), (\ref{Useries3d}), and (\ref{Epath3d}).}
\label{3Dalpha}
\end{center}
\end{figure}

\begin{figure}
\begin{center}
\includegraphics[scale=0.80]{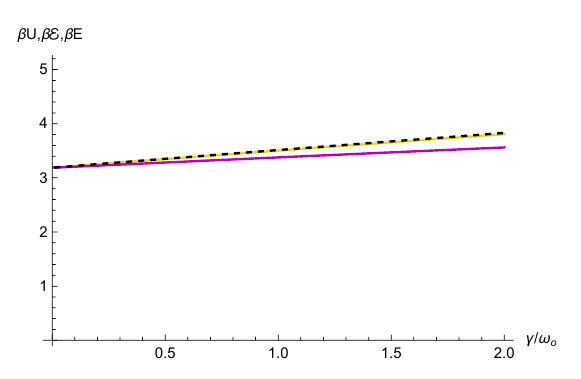}
\caption{Dimensionless energy functions \(\beta U\) (black-dashed), \(\beta \mathcal{E}\) (yellow-solid), and \(\beta E\) (violet-solid) for the dissipative magneto-oscillator, as a function \(\gamma/\omega_0\), for \(\alpha = \beta \hbar \omega_0 = 0.5\), \(\omega_c = 2.5 \omega_0\), and \(\omega_{\rm cut} = 10 \omega_0\). We have taken \(\omega_0 = 1\) and have considered the first 10,000 terms of Eqs. (\ref{E3dSeries}), (\ref{Useries3d}), and (\ref{Epath3d}).}
\label{3Dgamma}
\end{center}
\end{figure}

\section{Discussion}\label{DSec}
In this paper, we have discussed the various energy functions of the dissipative oscillator, also referred to as the Brownian oscillator. We discussed three unequal notions of thermal energy associated with the system. These were the average energy function \(E(T)\) as obtained via the fluctuation-dissipation theorem as the average of \(H_S\) over the noise ensemble; the energy function \(\mathcal{E}(T) = -\frac{\partial}{\partial \beta} \ln \mathcal{Z}\), where \(\mathcal{Z}\) is the reduced partition function; and finally \(U(T) = \langle H \rangle_{\rho_H} - \langle H_B \rangle_{\rho_{H_B}}\), which was termed the internal energy function. Quite remarkably, they were all found to agree for the simple case of Ohmic dissipation, leading to memoryless friction. The introduction of a memory timescale \(\omega_{\rm cut}^{-1}\) for dissipation renders the energy functions \(E(T)\) and \(U(T)\) unequal. However, the energy function \(\mathcal{E}(T)\) computed directly from the partition function is found to coincide with \(U(T)\) for the one-dimensional case,  while it slightly differs from the internal energy function for the three-dimensional case when a magnetic field is applied. \\

Let us now revisit the weak-coupling limit. As it turns out, in the weak-coupling limit, \(E(T)\), \(\mathcal{E}(T)\), and \(U(T)\) become equal to the thermally-averaged energy of a quantum harmonic oscillator in the canonical ensemble, a problem which is frequently described in various textbooks of statistical mechanics \cite{R}, especially in the context of an Einstein solid. This is rather expected, as the elementary textbook approach to deriving the canonical partition function involves the assumption that the system and the bath are weakly coupled. In a sense therefore, the functions \(E(T)\), \(\mathcal{E}(T)\), and \(U(T)\) are appropriate generalizations of the textbook result for the strong-coupling regime. Physically, in the weak-coupling limit, and up to a first approximation, only those oscillators of the heat bath whose frequencies are in resonance with the characteristic frequency of the system given by \(\omega_0\) can be considered relevant. This means we may drop the off-resonant frequencies, and approximately write Eq. (\ref{symmetricnoisecorrelation1}) as
\begin{equation}\label{ffff}
\langle \{ F(t), F(t') \} \rangle \approx \bigg[ 2 m \gamma \hbar \omega_0 \coth \bigg(\frac{\hbar \omega_0}{2k_B T}\bigg) \bigg] \delta(t-t').
\end{equation} Thus, one may define \cite{GSA}: \(\Gamma =m \gamma \hbar \omega_0 \coth \big(\frac{\hbar \omega_0}{2k_B T}\big)\), which, for \(\hbar \rightarrow 0\) appropriately reduces to \(\Gamma \rightarrow 2 m \gamma k_B T\), familiar from the classical Brownian motion problem.\\

We end by noting that in the classical limit, \(E(T)\), \(\mathcal{E}(T)\), and \(U(T)\) are all just \(k_BT\) (in one dimension) and requires no additional assumption of weak-coupling. This establishes the robustness of the classical equipartition theorem: at thermal equilibrium, the energy per quadratic term \footnote{Plus no coupling between any pair of variables, i.e., quadratic terms proportional to \(x^2\) and \(y^2\) are allowed, but \(xy\) is not.} in the Hamiltonian \(H_S\) is \(k_B T/2\), independent of the strength of coupling between the system and the heat bath. The same is true even in three dimensions with an applied magnetic field, implying that although an applied magnetic field may influence the energetics in the quantum mechanical case, in the classical limit, the energetics are independent of the applied magnetic field. This is, of course, quite well known as a consequence of the Bohr-van Leeuwen theorem.\\

\textbf{Acknowledements:} The authors dedicate this paper to Prof Sushanta Dattagupta on the occasion of his 75th birthday with deep respect and admiration. A.G. is grateful to Shamik Gupta for useful correspondences, and to the Ministry of Education (MoE), Government of India, for financial support in the form of a PMRF (ID: 1200454). J.K. gratefully acknowledges the financial support received from IIT Bhubaneswar, in the form of an Institute Research Fellowship. M.B. is supported by the Science and Engineering Research Board (SERB), Government of India, under the Core grant (Project No. CRG/2020/001768) and MATRICS grant (Project no. MTR/2021/000566).

\appendix

\section{Quantum Langevin equation for the dissipative oscillator}\label{PreSec}
For the model at hand, the heat bath is described by the following Hamiltonian:
\begin{equation}\label{HB}
H_B = \sum_{j=1}^N \Bigg[ \frac{p_j^2}{2m_j} + \frac{m_j \omega_j^2q_j ^2}{2} \Bigg],
\end{equation} where it will be understood that \(N \rightarrow \infty\). The heat bath is at a thermal equilibrium state with temperature \(T\), and in the absence of the system, its equilibrium distribution is just
\begin{equation}\label{rhoB}
\rho_{H_B} = \frac{e^{-H_B/k_B T}}{Z_B}.
\end{equation}
Now, starting with Eq. (\ref{Hmodel}), one can derive the Heisenberg equations for the system and bath variables, which read
\begin{eqnarray}
\frac{dx}{dt}&=&\frac{p}{m}, \hspace{2mm} \frac{dp}{dt}=-m \omega_0^2 x +\sum_{j=1}^N m_j \omega_j^2 \big(q_j-x\big), \label{Pdot}\\
\frac{dq_j}{dt}&=&\frac{p_j}{m_j}, \hspace{2mm} \frac{dp_j}{dt}=-m_j\omega_j^2\big(q_j-x\big) \label{pjEq.}.
\end{eqnarray}
 If one solves the equations of motion for the bath variables, i.e., Eq. (\ref{pjEq.}), and then substitutes them into those of the system, i.e., Eq. (\ref{Pdot}), one obtains the quantum Langevin equation as given in Eq. (\ref{qle}). In Eq. (\ref{qle}), the function \(\mu(t)\) is the dissipation kernel, which reads
\begin{equation}\label{1}
\mu(t) = \Theta(t) \sum_{j = 1}^N m_j \omega_j^2 \cos (\omega_j t) ,
\end{equation} where the presence of \(\Theta(t)\) ensures that \(\mu(t)\) vanishes for negative arguments, consistent with the principle of causality. In deriving Eq. (\ref{qle}), one assumes that the system and the bath are in a state of thermal equilibrium at the initial instant \cite{bez,physicaA}. Then, \(F(t)\) is a Gaussian noise whose expression reads
 \begin{equation}\label{noise}
 F(t)=\sum_{j=1}^{N} m_j \omega_j^2 \Bigg[\big(q_j(0)-x(0)\big)\cos(\omega_jt)+\frac{p_j(0)}{m_j\omega_j}\sin(\omega_jt)\Bigg].
 \end{equation}
Notice that the noise depends upon the initial position of the system as well as the initial conditions of the bath oscillators, which can be taken to be distributed according to the canonical distribution, i.e.,
\begin{equation}\label{p0}
\mathcal{P}(0) = \frac{\exp\Bigg\{{-\beta \sum_{j=1}^N \bigg[ \frac{p_j^2(0)}{2m_j} + \frac{m_j \omega_j^2 \big(q_j(0) - x(0) \big)^2}{2} \bigg]}\Bigg\}}{\Lambda},
\end{equation} where \(\Lambda\) is some normalizing factor. Eq. (\ref{p0}) tells us that we have taken the system and the heat bath to be at thermal equilibrium at the initial instant, i.e., at \(t = 0\). Then, with respect to this distribution, the noise operator \(F(t)\) is Gaussian, i.e., it has zero mean, all odd moments vanish, while the even moments can be written as ordered products of second moments. Since the noise explicitly depends upon the initial conditions, which can be chosen from the distribution given in Eq. (\ref{p0}), this defines an ensemble of noises, all characterized by different choices of the initial conditions. The spectral properties of the noise are given by the following correlation function and commutator \cite{purisdgbook}:
\begin{eqnarray}
\langle \lbrace F(t), F(t') \rbrace \rangle &=& \frac{2}{\pi}\int_{0}^{\infty}d\omega \hbar \omega {\rm Re} [ \tilde{\mu} (\omega)] \coth\Big(\frac{\hbar\omega}{2k_BT}\Big) \nonumber \\
&& \times \cos \lbrack \omega(t-t')\rbrack,  \label{symmetricnoisecorrelation1} \\
\langle \lbrack F(t), F(t') \rbrack \rangle &=& \frac{2}{i\pi}\int_{0}^{\infty}d\omega \hbar \omega   {\rm Re}[ \tilde{\mu} (\omega)] \sin\lbrack \omega(t-t')\rbrack , \nonumber \\
\label{noisecommutator1}
\end{eqnarray} where \(\tilde{\mu}(\omega)\) is the Fourier transform of \(\mu(t)\). \\

Therefore, in a sense, the choice of \(\tilde{\mu}(\omega)\) describes the details of the heat bath, i.e., it not only characterizes the dissipation term \(\int_{-\infty}^t \mu(t - t') \dot{x}(t') dt' \) of the quantum Langevin equation, but also the noise term \(F(t)\), via Eqs. (\ref{symmetricnoisecorrelation1}) and (\ref{noisecommutator1}). In this paper, we refer to the following choices of \(\tilde{\mu}(\omega)\) (here, \(\gamma, \omega_{\rm cut} > 0\)):
\begin{eqnarray}
\tilde{\mu}(\omega) &=& m \gamma, \\
 \tilde{\mu}(\omega) &=& \frac{m \gamma \omega_{\rm cut}^2}{\omega^2 + \omega_{\rm cut}^2} + i \frac{m \gamma \omega \omega_{\rm cut}}{\omega^2 + \omega_{\rm cut}^2},
\end{eqnarray} respectively called the Ohmic and Drude models. As can be verified by taking an inverse Fourier transform, these two cases correspond respectively to friction kernels: \(\mu(t) \sim \delta(t)\) (Ohmic) and \(\mu(t) \sim e^{- \omega_{\rm cut}t}\) (Drude). Therefore, Ohmic dissipation is memoryless, in the sense that the dissipation force is instantaneous, while Drude dissipation has exponentially decaying memory. The parameter \(\omega_{\rm cut}\) characterizing the memory of the bath is called the Drude cutoff and for \(\omega_{\rm cut} \rightarrow \infty\), Drude dissipation is approximately Ohmic. We note that the real and imaginary parts of \(\tilde{\mu}(\omega)\) are respectively, even and odd (see also \cite{QLE}).

\section{Positivity and normalization of \(P_E(\omega)\)}\label{appendixA}
In this appendix, we outline a proof of the normalization of \(P_E(\omega)\) for the one-dimensional dissipative oscillator (see \cite{jarzy1,jarzy2,jarzy3,jarzy4,jarzy5,kaur} for similar calculations) and also show that it is positive definite. The case of the three-dimensional magneto-oscillator has been treated in \cite{kaur}. If \(P_E(\omega)\) is normalized, we could conclude that \(P_E(\omega)\) is a genuine probability distribution function, provided that we can ensure that \(P_E(\omega) \geq 0\), \(\omega \in [0,\infty)\). The positivity can be verified very straightforwardly, because from Eq. (\ref{alphadef11}), we have
\begin{equation}
{\rm Im} [\alpha^{(0)}(\omega)] = \frac{1}{m} \frac{\omega {\rm Re} [\tilde{\gamma}(\omega)]}{\big(\omega_0^2 - \omega^2 + \omega {\rm Im} [\tilde{\gamma}(\omega)]\big)^2 + (\omega {\rm Re} [\tilde{\gamma} (\omega)])^2},
\end{equation} where \(\tilde{\gamma}(\omega) = \tilde{\mu}(\omega)/m\). Now since \({\rm Re} [\tilde{\gamma}(\omega)] > 0\), on the real line \cite{QLE}, a requirement emerging from the second law of thermodynamics, one concludes that \(P_E(\omega) \geq 0\), for  \(\omega \in [0,\infty)\). Next, let us consider the issue of verifying normalization. The expression for \(P_E(\omega)\), as given in Eq. (\ref{P2generic}) can be expressed as
\begin{equation}\label{PESum}
P_E(\omega) = \frac{P_k(\omega) + P_p(\omega)}{2},
\end{equation} where
\begin{equation}
P_k(\omega) = \frac{2m\omega }{\pi}{\rm Im} [\alpha^{(0)}(\omega + i 0^+)],
\end{equation}
\begin{equation}\label{pp}
P_p(\omega) = \frac{2m\omega_0^2 }{\pi \omega}{\rm Im} [\alpha^{(0)}(\omega + i 0^+)],
\end{equation} respectively, corresponding to the kinetic and potential energy parts. Thus, if one could show that \(P_k(\omega)\) and \(P_p(\omega)\) are normalized individually, then from Eq. (\ref{PESum}), we may conclude that \(P_E(\omega)\) is normalized. 

\subsection{Normalization of \(P_k(\omega)\)}
To prove that $P_k(\omega)$ is normalized, let us first recall the following two properties of the Laplace transform operator:
\begin{eqnarray}
\mathcal{L} [\dot{x}(t)] &=& s \mathcal{L} [x(t)] - x(0), \\
\mathcal{L} [\ddot{x}(t)] &=& s^2 \mathcal{L} [x(t)] - s x(0) - \dot{x}(0).
\end{eqnarray}

Denoting \(\mathcal{L} [x(t)] = \hat{x}(s)\), the quantum Langevin equation gives
\begin{equation}\label{langeqnlaplaced}
(ms^2  + s \hat{\gamma}(s) + m \omega_0^2) \hat{x}(s) = m \dot{x}(0) + m x(0) + \hat{f}(s).
\end{equation}
We now denote
\begin{eqnarray}
\mathcal{L}[Q(t)] &=& \hat{Q}(s) = \frac{1}{ms^2 + s \hat{\mu}(s) + m \omega_0^2}, \\
 \mathcal{L}[R(t)] &=& \hat{R}(s) = \frac{ms}{ms^2 + s \hat{\mu}(s) + m \omega_0^2},
\end{eqnarray}
and then, Eq.~(\ref{langeqnlaplaced}) can be expressed as
\begin{equation}
\hat{x}(s) = m \dot{x}(0) \hat{Q}(s) + x(0) \hat{R}(s) + \hat{Q}(s) \hat{f}(s).
\end{equation}
Thus, we have
\begin{equation}
{x}(t) = p(0) Q(t) + x(0) R(t) + \int_0^t dz~Q(t-z) f(z),
\end{equation} where we have put \(p(0) = m \dot{x}(0)\). Working along similar lines, one can show that the momentum operator can be expressed as
\begin{equation}\label{momentumforlaplace}
p(t) = p(0) R(t) + m x(0) \dot{R}(t) + \int_0^t dz~R(t-z) f(z).
\end{equation}
\begin{widetext}
Equation~(\ref{momentumforlaplace}) will allow us to determine the mean kinetic energy of the system, via the following correlation function:
\begin{equation}
\langle \{p(t), p(t')\} \rangle = \int_0^t dt_1 \int_0^{t'} dt_2~R(t - t_1) R(t' - t_2) \langle \{f(t_1), f(t_2)\} \rangle.
\end{equation}
Now, using the fluctuation-dissipation theorem, we have
\begin{equation}
\langle \{p(t), p(t')\} \rangle = \int_0^\infty d\omega~ \widetilde{C}_{ff}(\omega) \int_0^t dt_1 \int_0^{t'} dt_2~R(t - t_1) R(t' - t_2) \cos [\omega(t_1-t_2)],
\end{equation} where \(\widetilde{C}_{ff}(\omega) = (\hbar \omega/2) \coth \big(\hbar \omega/(2 k_B T)\big) \mathrm{Re}[\widetilde{\mu}(\omega)\)] is the cosine transform of the noise autocorrelation function. Next, we put \(t = t'\), giving us
\begin{equation}
\langle p^2(t) \rangle = \int_0^\infty d\omega~\widetilde{C}_{ff}(\omega) \int_0^t d\tau \int_0^{t} du~R(\tau) R(u) \cos [\omega(\tau-u)],
\end{equation} where we have put \(\tau = t - t_1\) and \(u = t - t_2\).
\end{widetext}
For the steady state, we take the limit \(t \rightarrow \infty\), which gives
\begin{equation}\label{Ektgoestoinfty}
E_k = \lim_{t \rightarrow \infty} \frac{\langle p(t)^2 \rangle}{2m} = \frac{1}{2m} \int_0^\infty d\omega~\widetilde{C}_{ff}(\omega) I(\omega),
\end{equation} with
\begin{eqnarray}
I(\omega) &=& \int_0^\infty d\tau \int_0^{\infty} du~R(\tau) R(u) \cos [\omega(\tau-u)] \nonumber \\
&=& \frac{1}{2} \int_0^\infty d\tau~R(\tau) e^{i \omega \tau} \int_0^\infty du~R(u) e^{- i \omega u} \nonumber \\
&&+ \frac{1}{2} \int_0^\infty d\tau~ R(\tau) e^{-i \omega \tau} \int_0^\infty du~R(u) e^{ i \omega u} \nonumber \\
&=& \hat{R}(i \omega) \hat{R} (-i\omega).
\end{eqnarray}
Substituting this into Eq.~(\ref{Ektgoestoinfty}), it follows that
\begin{equation}
E_k = \frac{1}{2m} \int_0^\infty d\omega~ \widetilde{C}_{ff}(\omega)  \hat{R}(i \omega) \hat{R} (-i\omega).
\end{equation}
This means one identifies
\begin{eqnarray}
P_k(\omega) &=& \frac{1}{m} {\rm Re} [\widetilde{\mu}(\omega)] \hat{R}(i \omega) \hat{R} (-i\omega) \nonumber \\
&=& [\hat{R}(i \omega) + \hat{R} (-i\omega)] := R_F (\omega).
\end{eqnarray}
It turns out that \(R_F(\omega)\) is the cosine transform of \(R(t)\), i.e.,
\begin{equation}
R_F(\omega) = \frac{2}{\pi} \int_0^\infty dt~R(t) \cos (\omega t),
\end{equation} therefore implying
\begin{equation}
R(t) = \int_0^\infty d\omega~R_F(\omega) \cos (\omega t).
\end{equation}
Putting \(t = 0\), we find
\begin{equation}
R(0) = \int_0^\infty d\omega~R_F(\omega).
\end{equation}
We now use the initial-value theorem of the Laplace transform to give
\begin{equation}
R(0) = \lim_{s \rightarrow \infty}  s \hat{R}(s) = 1,
\end{equation}which means
\begin{equation}
 \int_0^\infty d\omega~R_F(\omega)=  \int_0^\infty d\omega~P_k(\omega)= 1.
\end{equation}
This proves that \(P_k(\omega)\) is normalized. 

\subsection{Normalization of \(P_p(\omega)\)}
We will now prove that the function \(P_p(\omega)\) in Eq.~(\ref{pp}) is normalized. We consider the following integral:
\begin{equation}\label{normalcondpp}
\int_0^\infty d\omega~P_p (\omega)= \frac{2 m \omega_0^2}{\pi} \int_0^\infty d\omega~\frac{{\rm Im} [\alpha^{(0)} (\omega)] }{\omega}.
\end{equation}
Next, we recall the following relationship from \cite{landau}:
\begin{equation}
\alpha^{(0)} (i \omega) = \frac{2}{\pi} \int_0^\infty ds~\frac{{\rm Im} [\alpha^{(0)} (s)] }{\omega^2 + s^2} s,
\end{equation} which means putting \(\omega =0\) gives us
\begin{equation}\label{normalpp2}
\alpha^{(0)} (0) = \frac{2}{\pi} \int_0^\infty d\omega~\frac{{\rm Im} [\alpha^{(0)} (s)] }{s}.
\end{equation} Therefore, if we combine Eqs.~(\ref{normalcondpp}) and~(\ref{normalpp2}), we obtain
\begin{equation}
\int_0^\infty d\omega~P_p (\omega)= m \omega_0^2 \alpha^{(0)}(0).
\end{equation}
From Eq. (\ref{pp}), it directly follows that putting \(\omega =0\) gives \(\alpha^{(0)}(0) = [m \omega_0^2]^{-1}\) which means \(P_p(\omega)\) is normalized. Since \(P_k(\omega)\) and \(P_p(\omega)\) are individually normalized, we conclude from Eq. (\ref{PESum}) that \(P_E(\omega)\) is normalized over \(\omega \in [0,\infty)\). Thus, \(P_E(\omega)\) is a genuine probability distribution function as has been considered in this paper.

\section{Positivity and normalization of \(P_U(\omega)\)}\label{appendixB}
In this appendix, we outline a proof of the normalization of \(P_U(\omega)\) for the one-dimensional dissipative oscillator. The case of the three-dimensional magneto-oscillator has been treated in \cite{kaur2}. We recall that the susceptibility can be expressed as \cite{FordQT0}
\begin{equation}
\alpha^{(0)}(\omega) = -\frac{1}{m} \frac{\prod_{j=1}^N (\omega^2 - \omega_j^2)}{\prod_{k=0}^N (\omega^2 - \Omega_k^2)},
\end{equation} where \(\{\omega_j\}\) are the frequencies of the heat-bath oscillators (for the isolated heat bath), while \(\{\Omega_k\}\) are the normal-mode frequencies including the dissipative oscillator, while it is interacting with the heat bath.

\begin{widetext}
From the well-known formula $1/(x+i0^+)=\mathrm{P}(1/x)-i\pi\delta(x)$, with $\mathrm{P}$ denoting the principal part, it follows that \(P_U(\omega)\) now reads
\begin{eqnarray}\label{fnormaltheorem}
P_U(\omega)&=&\frac{1}{\pi}\mathrm{ Im}\Bigg\lbrack\frac{\mathrm{d}}{\mathrm{d}\omega}\ln\alpha^{(0)}(\omega)\Bigg\rbrack\nonumber \\
&=& \sum_{k=0}^N \big[ \delta (\omega - \Omega_k) + \delta (\omega + \Omega_k) \big]  -  \sum_{j=1}^N \big[ \delta (\omega - \omega_j) + \delta (\omega + \omega_j) \big].
\end{eqnarray}
\end{widetext}
Thus, the function \(P_U(\omega)\) is positive definite, and is also normalized, because
\begin{equation}
\int_0^\infty P_U(\omega) d\omega = (N+1) - N = 1.
\end{equation}
This means the function \(P_U(\omega)\) is a genuine probability distribution function, as has been treated in this paper. We should however remember that except for the case of Ohmic dissipation, the functions \(P_E(\omega)\) and \(P_U(\omega)\) do not coincide.

\end{document}